\documentclass[letterpaper]{article} 
\usepackage{aaai2026}  
\usepackage{times}  
\usepackage{helvet}  
\usepackage{courier}  
\usepackage[hyphens]{url}  
\usepackage{graphicx} 
\usepackage{natbib}  
\usepackage{caption} 
\usepackage{algorithm}
\usepackage{algorithmic}
\usepackage{xcolor}
\newcommand{\answerYes}[1]{\textcolor{blue}{#1}} 
 
\newcommand{\answerNA}[1]{\textcolor{gray}{#1}} 
 
\usepackage{newfloat}
\usepackage{listings}
\DeclareCaptionStyle{ruled}{labelfont=normalfont,labelsep=colon,strut=off} 
\floatstyle{ruled}
\newfloat{listing}{tb}{lst}{}
\floatname{listing}{Listing}

\usepackage{tablefootnote}
\usepackage{booktabs}
\title{The Great Data Standoff: Researchers vs. Platforms Under the Digital Services Act}
\author{
    Catalina Goanta\textsuperscript{\rm 1},
    Savvas Zannettou\textsuperscript{\rm 2},
    Rishabh Kaushal\textsuperscript{\rm 3},
    Jacob van de Kerkhof\textsuperscript{\rm 4}\footnote{Work completed while employed by Utrecht University},
    Thales Bertaglia\textsuperscript{\rm 1},
    Taylor Annabell\textsuperscript{\rm 5},
    Haoyang Gui\textsuperscript{\rm 1},
    Gerasimos Spanakis\textsuperscript{\rm 6},
    Adriana Iamnitchi\textsuperscript{\rm 6}
}
\affiliations{
    \textsuperscript{\rm 1}Utrecht University\\
    \textsuperscript{\rm 2}TU Delft\\
    \textsuperscript{\rm 3}Indira Gandhi Delhi Technical University for Women\\
    \textsuperscript{\rm 4}De Brauw Blackstone Westbroek\\
    \textsuperscript{\rm 5}Cardiff University\\
    \textsuperscript{\rm 6}Maastricht University\\
    e.c.goanta@uu.nl, s.zannettou@tudelft.nl, rishabhkaushal@igdtuw.ac.in, j.j.w.vandekerkhof@uu.nl, t.f.costabertaglia@uu.nl, annabellt@cardiff.ac.uk, h.gui@uu.nl, jerry.spanakis@maastrichtuniversity.nl, a.iamnitchi@maastrichtuniversity.nl
}

\usepackage{bibentry}

\begin{document}

\maketitle

\begin{abstract}
    
To facilitate accountability and transparency, the Digital Services Act (DSA) sets up a process through which Very Large Online Platforms (VLOPs) need to grant vetted researchers access to their internal data (Article 40(4)). Although expectations are high for this mechanism to provide unprecedented data access as part of the DSA compliance obligations, operationalising it is challenging for two reasons. First, data access is only available for research on systemic risks affecting European citizens, a concept that still raises high levels of legal uncertainty. Second, data access suffers from an inherent standoff problem. Researchers need to request specific data but are not in a position to know all internal data collected, processed and stored by VLOPs, who, in turn, expect and demand data specificity for potential access. To contribute to the discussion of how Article 40 can be interpreted and applied, we provide a concrete illustration of what data access can look like in a real-world systemic risk case study. Our goal is not only to share with the research community at large a scenario for reflection, but also to provide hands-on insights into what type of data platforms may be required to share through the DSA. To this end, we focus on the 2024 Romanian presidential election interference incident, as this event is the first of its kind to trigger systemic risk investigations by the European Commission. In the context of these elections, one candidate is said to have benefited from TikTok algorithmic amplification through a complex and multilayered dis- and misinformation campaign. By analysing this incident, we can concretely comprehend election-related systemic risk in order to explore practical research tasks and compare necessary data with available TikTok data. In particular, our study makes two contributions: (i)~we combine insights from law, computer science and platform governance to shed light on the complexities of studying systemic risks in the context of election interference, focusing on two relevant factors: platform manipulation and hidden advertising; and (ii)~we provide practical insights into various categories of available data for the study of TikTok, based on platform documentation, data donations and TikTok's Research API.

\end{abstract}
\maketitle

\section{Introduction}
The Digital Services Act (DSA) has introduced new transparency obligations for Very Large Online Platforms (VLOPs) offering their services to European Union (EU) users. One of these obligations is to grant access to their internal data to researchers through a procedure established by the DSA (Article 40) and further detailed in the accompanying Delegated Act (DA). While Article 40 promises to provide unprecedented data access, two main obstacles hinder operationalisation. First, data access is only granted for the study of systemic risks (Article 40(4) DSA). Systemic risks are a new concept in European platform regulation, and although they are somewhat defined in the DSA (Article 34), a lot of uncertainty remains regarding their practical interpretation. 
Second, VLOPs are opaque organisations, so there is no way of knowing what data they actually gather, infer, process, and use. However, researchers will be expected to specify the data they require. Not having a full picture of the data VLOPs have can render data access less effective than intended. In light of these major interpretational and operational limitations, data access under Article 40 DSA remains, for the time being and most part, shrouded in mystery.

To contribute to the discussion of how Article 40 can be interpreted and applied, we provide a concrete illustration of what data access can look like in a real-world systemic risk case study. Our goal is not only to share with the research community at large a scenario for reflection, but also to provide hands-on insights into what type of data platforms may be required to share through the DSA. To this end, we focus on the 2024 Romanian presidential election interference incident~\cite{ross_georgescu_2024} as this event is the first of its kind to trigger systemic risk investigations by the European Commission \cite{eu_commission_2024}. In the context of these elections, one candidate is said to have benefited from TikTok algorithmic amplification through a complex and multilayered coordinated dis- and misinformation campaign. By analysing this incident, we explore practical research tasks and compare necessary data with available TikTok data. In particular, our study makes two contributions: (i)~we combine insights from law, computer science, and platform governance to highlight the complexities of studying systemic risks in the context of election interference, focusing on two relevant factors: platform manipulation and hidden advertising; and (ii)~we provide comprehensive and practical insights into various categories of available data for the study of TikTok, based on platform documentation, data donations, and TikTok's Research API.


\section{Data Access and Systemic Risks Under the Digital Services Act}
\label{sec:DSA}

Article 40 DSA is a novel transparency instrument, seeking to build on existing measures such as transparency reporting~\cite{urman_how_2023, kosta_government_2019}, statements of reasons~\cite{leerssen_end_2023} and the DSA transparency database~\cite{kaushal_automated_2024}. It gives Digital Services Coordinators (DSCs, Article 40(1) DSA) and ‘vetted researchers’ (Article 40(8) DSA) unprecedented data access to online platforms to unpack opaque algorithmic systems~\cite{leerssen_outside_2024, pasquale_black_2016}. Particularly in a landscape where platforms (including social media platforms) are tightening their control over data and deprecating APIs,\footnote{For example, in 2024 Meta discontinued CrowdTangle: \url{https://transparency.meta.com/en-gb/researchtools/other-datasets/crowdtangle/.}} Article 40 DSA makes a lofty promise. However, two main considerations severely restrict what may otherwise be seen as a new tool for wide platform transparency. 

First, the procedure to acquire data access through Article 40 is complex. It is outlined in the DA, an instrument adopted by the European Commission to further clarify who can apply for data access, and under which circumstances. Interested researchers must submit an application that contains, amongst others, a description of the research project, the research question, the systemic risks or mitigation measures studied, and the planned research activities (Article 8 DA) with the DSC of their Member State, or with the DSC of the Member State of establishment of the platform. Requests may be submitted both at the Member State of establishment of the researcher and that of the platform, but the former has to forward the request to the latter: the DSC of the Member State of establishment of the platform is the only authority with jurisdiction in these cases. After submission of the data access application, the DSC of the Member State of establishment of the platform formulates a reasoned request, in which it determines the access modalities -- the avenue through which the data provider grants data -- and forwards the application to the online platform (Article 9 and 10 DA). The platform then decides whether it grants the request or submits amendments to the DSC (Article 40(5) DSA). After an optional mediation period (Article 13 DA), the data access request is completed and access should be granted. The role of the DSC in this process is crucial, as it defines the final request submitted to the platform. As such, the likelihood of a successful data application can be contingent on aligning that application with the DSC's interests, and whether the DSC has the operational capacity to address the application. It is likely that the DSCs of establishment of most VLOPs (Ireland and the Netherlands) will be subject to most applications, which could lead to a case overload paralysing public authorities.

Second, data access is conditional on making a contribution to the detection, identification, and understanding of systemic risks in the EU or assessing whether those risks are sufficiently mitigated (Article 40(4) DSA). This makes systemic risks central to data access requests. The concept leaves room for a lot of interpretation. The DSA does not provide a clear definition, but shapes the meaning of systemic risks in Article 34(1). The article uses four \textit{non-exhaustive}~\cite{hofmann_digital_2025} examples to illustrate systemic risks: (a)~the dissemination of illegal content; (b)~actual or foreseeable negative effects for the exercise of fundamental rights; (c)~actual or foreseeable negative effects on civic discourse and electoral processes and public security; (d)~actual or foreseeable negative effects in relation to gender-based violence, the protection of public health and serious negative consequences to the person’s physical and mental well-being. 
In assessing these risks, online platforms can include contributing factors listed in the second paragraph of the article: (i)~the design of their recommender system and other algorithmic systems; (ii)~their content moderation systems; (iii)~the applicable terms and conditions and their enforcement; (iv)~systems for selecting and presenting advertisements; and (v)~data related practices of the provider. The risks may also be influenced by intentional manipulation of the platform service, for example, through inauthentic use or the possible virality of illegal or harmful content content.

As systemic risks are so broad, this can create challenges for data access requests. For example, if a researcher was interested in studying recommender systems that amplify dissemination of hate speech~\cite{weimann_research_2023}, this can fall under the ‘dissemination of illegal content’ category since certain types of hate speech are illegal~\cite{hietanen_towards_2023}.\footnote{See also Council Framework Decision 2008/913/JHA of 28 November 2008 on combating certain forms and expressions of racism and xenophobia by means of criminal law; European Court of Human Rights, \textit{Erbakan v Turkey}, para 56.} However, hate speech can have negative effects on the exercise of fundamental rights, such as human dignity, private life, and non-discrimination, which are all well established in European Court of Human Rights case law.\footnote{See for instance European Court of Human Rights, \textit{Lenis v Greece}, para 55; \textit{Beizaras \& Lecikas v Lithuania}, para 117; \textit{Ivanov v Russia}.} It can even have negative effects on civic discourse, elections, and public security.\footnote{See for instance European Court of Human Rights, \textit{Sanchez v France}.} Moreover, hate speech can have serious consequences for gender-based violence and a person’s mental well-being.\footnote{See for instance European Court of Human Rights \textit{Beizaras \& Lecikas v Lithuania}.} It is therefore hard to categorise the platform features leading to systemic risks. This represents an additional barrier in how researchers frame their data access requests.

\section{Data Needed for Systemic Risks Research: Election Interference as a Case Study}
\label{sec:ro-elections}

Election interference has been one of the original smoking guns in terms of abusing social media architectures at scale to benefit private interests at the detriment of democratic processes. A well-known example in this respect is the Cambridge Analytica scandal~\cite{meredith_facebook-cambridge_2018}, where private actors accessed Facebook data to profile and target voters in the United States 2016 presidential election and in the Brexit referendum~\cite{guardian2018}. The vulnerabilities of social media platforms, particularly during elections, have also come to light in the recent Romanian presidential elections of 2024 \cite{lutac_reteta_2024}.

We take this incident as a case study to concretely discuss what data could be requested and for what systemic risks under the DSA's data access regime for vetted researchers. This case study is important for at least two reasons. First, the Romanian situation is the first post-DSA election interference investigation by the European Commission~\cite{eu_commission_2024}. This raises urgent practical questions such as what precisely needs to be studied in this context and, most importantly, what data is necessary to investigate systemic risks and mitigation strategies to illustrate what barriers researchers might face in requesting access to data under the DSA. Second, this case study contributes to existing scholarly debates around election interference, which have traditionally been US-centric or focused on Western Europe \cite{chong2020, eli2020}, whereas Eastern Europe still maintains significant understudied post-communist complexities. 
To operationalise these implications for data access, we first outline the known features of the alleged Romanian election interference. Second, we introduce two factors that we argue lead to a systemic risk (platform manipulation and hidden advertising) and discuss them by extracting data needs from relevant research studies and further proposing specific computational research tasks as examples.

\subsection{A Brief Overview of an Election Case Study}

The Romanian presidential election campaign took place between October 25 and November 24, 2024. According to investigative journalists and Romanian intelligence agencies~\cite{lutac_reteta_2024, snoop2025, presidency2025}, TikTok is the platform most heavily exploited to influence voters to support a right-wing candidate who went in a couple of weeks from being unknown to the most popular candidate in the voting results of the first voting round on November 24, 2024. The candidate's engagement relied on positive comments driving engagement and video views on TikTok. The amplified content included, among others, a combination of conspiracy theories, the rejection of Western interpretations of fundamental rights perceived as overly liberal, fascism, misogyny, and racism. This was also reflected by the artificial and later organic engagement involving this content~\cite{snoop2025}. Overall, the amplification also triggered TikTok's search recommendations, a feature unique to TikTok's increasing use as a search engine, to curate, deliver and further amplify content from and about the candidate. Suspecting manipulation and citing possible foreign state interference, Romania's Constitutional Court annulled the results of the first round of the presidential elections~\cite{ccr2024}, and the European Court of Human Rights refused the candidate's request for an interim measure to suspend the Constitutional Court's decision~\cite{echr2025}.

Preliminary evidence~\cite{lutac_reteta_2024, snoop2025, presidency2025} points to the use of TikTok in the Romanian elections through a combination of: (i)~\textbf{Inauthentic behaviour}: A coordinated network of direct promotion accounts generated fake engagement and triggered TikTok's search recommendations; (ii)~\textbf{Political influencer marketing}: At least one non-disclosed micro-influencer campaign revolving around the promotion of an (unnamed) ideal candidate; and (iii)~\textbf{Livestream gifting}: TikTok's streaming monetisation affordances (coins and gifts) were used to amplify (undisclosed) political advertising and political content.

This preliminary evidence is important in establishing a general narrative around how systemic risks might have been shaped in this election interference incident. However, investigative journalists can only analyse TikTok data externally by scraping it themselves or by relying on third-party tools~\cite{lutac_reteta_2024}, and intelligence agencies are not clear about how exactly they investigated the platform~\cite{presidency2025}. DSA data access for academic researchers is thus an avenue to study systemic risks using scientific methodologies. Yet, in order to do so, researchers must first identify the actual systemic risks and the data they need to investigate. 

The systemic risks surrounding the alleged interference in the Romanian presidential elections are multidimensional. The most obvious link in this case is to the election-related systemic risk. However, the other mentioned categories of systemic risks are also relevant: the dissemination of hidden advertising or hate speech (illegal content) (Article 34(1)(a)), the prevalence of hate speech as fundamental rights violations (Article 34(1)(b)), or the promotion of conspiracy theories affecting public health (Article 34(1)(d)) can also be systemic risks associated with the Romanian situation. In addition, these risks stem from a multitude of factors, as listed in Article 34(2), such as the design of TikTok's recommender system, its moderation systems, its policies regarding advertisements, or even the overall manipulation of the platform itself. In the context of this case study, we propose that in understanding the systemic risk of election interference in Romania, as mentioned in Article 34(1)(c), two factors are central: platform manipulation (Section~\ref{sec:manipulation}) and hidden advertising (Section~\ref{sec:hidden-ads}). 

\subsection{Platform Manipulation and Election Interference: Related Studies and Data Needs}
\label{sec:manipulation}

Platform manipulation, as relevant to election manipulation, refers to the deliberate use of various social media platform affordances to influence public opinion, behaviour, or perception, often including the use of deceptive and unethical means~\cite{akhtar2024sok,martino2020survey}. This can involve the spread of false information on social media platforms~\cite{vosoughi2018spread,ferrara2015manipulation}, the use of automated accounts (bots) to amplify online content~\cite{woolley2016automating,bessi2016social}, and coordinated campaigns designed to mislead or sway public opinion~\cite{zannettou2019disinformation,zannettou2019let}.
It can also involve the deliberate manipulation of specific platform affordances such as recommendation algorithms in an attempt to amplify content favouring one perspective while suppressing content from another, usually conflicting viewpoint~\cite{santos2021link,huszar2022algorithmic}.
This section reviews some of the documented techniques used for trying to influence the outcome of elections with the objective of identifying the data necessary for empirical studies. 

\textbf{Identifying False Information During Elections.} The spread of false information on social media platforms represents a significant and persistent challenge in the digital age~\cite{zannettou2019web,ferrara2020characterizing}. The problem of false information arises from the ease with which online content can be created, shared, and amplified across a large number of users. 
False information, irrespective of whether intentional or unintentional, has the potential to influence public opinion and change voter behaviour, exacerbate online polarisation, and even affect the integrity of electoral processes.

In response to this issue, social media platforms employ content moderation strategies, encompassing diverse measures designed to remove false information, demote it, or leave it accessible while providing additional context to help users critically assess its veracity. For instance, during the 2020 US elections, Twitter employed warning labels to inform users about false information related to political narratives~\cite{zannettou2021won}.

Due to the potential impact of spreading misleading information, the research community has devoted significant resources to studying false information on social media platforms in various disciplines, with research focusing on exposure to untrustworthy websites~\cite{guess2020exposure}, analysing the spread of false information during elections and its impact on society~\cite{allcott2017social,bovet2019influence,grinberg2019fake} and designing mitigation strategies like the use of fact-checking~\cite{clayton2020real}, nudges~\cite{pennycook2020fighting} or warnings to end-users about the veracity of information~\cite{zannettou2021won}. 
Below, we summarise some notable work focusing on false information during elections and the data they used.

Allcott and Gentzkow~\cite{allcott2017social} studied false information on social media during the 2016 US elections. By combining web browsing data, user surveys, and data from fact-checkers, they empirically investigate the prevalence of false information during the 2016 US elections. This study used many diverse data fields: 1) \textit{Content Data} (browsing histories), 2) \textit{User Data} (from user surveys), 3) \textit{Engagement Data} (how many times each real/fake article was shared), and 4) \textit{Moderation-related Data} (i.e., whether the articles are determined as fake or real by fact checkers).
Grinberg et al.~\cite{grinberg2019fake} studied the spread of fake news on Twitter and its connection with voter behaviour during the 2016 US presidential elections. The authors analysed public voter registration records on Twitter accounts, focusing on the exposure and spread of false information. The study used: 1)~\textit{Content Data} (tweets content), 2)~\textit{User Data} (Twitter account data), 3)~\textit{Network Data} (follower network on Twitter), and 4) \textit{Temporal data} (fake news exposure over time).
Bovet and Makse~\cite{bovet2019influence} studied the influence of false information on Twitter by analysing a large-space dataset of tweets that included links to news outlets during the 2016 US elections. By characterising the networks of information flows, they identified the most important influencers for sharing false information and their impact on the US elections. The study used: 1)~\textit{Content Data} (tweets content), 2)~\textit{User Data} (Twitter account data), and 3)~\textit{Network Data} (follower network on Twitter).
Other research focused on analysing content moderation interventions on false information related to elections. Specifically, \citet{zannettou2021won} studied the application of soft moderation interventions (i.e., warning labels) during the 2020 US elections, finding, among other things, that tweets with warning labels received more engagement than ones without. This work used the following data: 1)~\textit{Content Data} (tweets content), 2)~\textit{User Data} (Twitter account data), 3)~\textit{Moderation Data} (which tweets received warning labels), and 4) \textit{Temporal Data} (Tweets with warning labels over time).

\textbf{Identifying Coordinated Campaigns.} Coordinated campaigns refer to organised efforts to amplify or push specific narratives on social media. The goals of these efforts vary from sowing public discord to achieving various political or commercial goals. Such campaigns usually employ multiple accounts that are operated by automated bots or humans to execute a unified strategy.

Coordinated campaigns often use automated accounts (bots). The detection of bots used for content amplification and algorithmic manipulation has been a significant research focus. The use of automated accounts requires the creation/purchasing of accounts that are posing as regular human accounts. These fake accounts are controlled by automated means and aim to assist in the spread or amplification of specific content. In many cases, such accounts are dormant for years and are activated for a particular purpose. 
The research community has studied automated accounts on social media during elections~\cite{ferrara2020characterizing}, focusing mainly on understanding their role and behaviour~\cite{shao2018spread,ferrara2017disinformation}
and assessing their impact~\cite{bessi2016social,eady2023exposure}.

Specifically, Shao et al.~\cite{shao2018spread} analyse social bots and their role in sharing low-credibility content, finding that during the 2016 US elections, social bots shared a disproportionate amount of low-credibility content. To conduct their study, they used the following data: 1)~\textit{Content Data} (tweets), 2)~User data (Twitter account information), 3)~\textit{Moderation Data} (news sources with low credibility information), and 4) \textit{Engagement Data} (retweet information).
Bessi and Ferrara~\cite{bessi2016social} investigate how social bots affect political discussion during the 2016 US elections, finding that the presence of bots can negatively affect political discussion on Twitter. To conduct their research, they relied on the following data: 1)~\textit{Content Data} (tweets), 2)~\textit{User Data} (Twitter account information), 3) \textit{Temporal Data} (tweets over time), and 4)~\textit{Geospatial Data} (location of the tweets).
The same kind of data was also applied by Ferrara~\cite{ferrara2017disinformation} to study social bots during the 2017 French election.
Ferrara et al.~\cite{ferrara2020characterizing} focus on the 2020 US elections and show that social bots online exacerbate the problem of political echo chambers. For their work, they used the following data: 1)~\textit{Content Data} (tweets), 2)~\textit{User Data} (Twitter account information), 3) \textit{Temporal Data} (tweets over time), 4)~\textit{Moderation Data} (tweets shared by state-sponsored actors), and 5) \textit{Engagement Data} (retweet information).

Another well-known implementation of such campaigns uses \textit{state-sponsored trolls}, which are humans who operate a set of fake accounts, are sponsored by their governments, and aim to push specific narratives on social media to push their government's agenda. Examples include Russian trolls active on social media platforms during the 2016 US elections~\cite{badawy2018analyzing}. 
Previous work on coordinated campaigns using state-sponsored trolls focused on analysing their characteristics and the strategies employed~\cite{zannettou2019disinformation}, detecting actors involved in such campaigns during elections~\cite{luceri2020detecting}, and understanding their impact and influence on the Web and our society~\cite{golovchenko2020cross,badawy2018analyzing}. 
Specifically, Badawy et al.~\cite{badawy2018analyzing} investigate the Russian interference campaign on Twitter during the 2016 US presidential election. They analysed a large dataset of tweets to demystify the strategies used by the Internet Research Agency (IRA), a Russian organization known for its online propaganda, and analysed the Twitter users that engaged with tweets shared by IRA. The data used for this study include: 1)~\textit{Content Data} (tweets), 2)~\textit{User Data} (Twitter account information), 3)~\textit{Network Data} (retweet network), 4)~\textit{Moderation Data} (Accounts identified as IRA accounts),
5)~\textit{Temporal Data} (IRA tweets over time), 6)~\textit{Engagement Data} (people who engaged with IRA tweets), and 7)~\textit{Geospatial Data} (Location of tweet posting).
Zannettou et al.~\cite{zannettou2019disinformation} analyse the behaviour and strategies employed by IRA trolls on Twitter during the 2016 US elections and assess the influence that they had on other platforms like 4chan and Reddit. This study leveraged the following data: 1)~\textit{Content Data} (tweets), 2)~\textit{User Data} (Twitter account information), 3)~\textit{Moderation Data} (Accounts identified as IRA accounts),
4)~\textit{Temporal Data} (IRA tweets over time), and 5)~\textit{Geospatial Data} (Location of tweet posting).
Similarly, Golovchenko et al.~\cite{golovchenko2020cross} investigated IRA's tweets during the 2016 US elections, focusing on analysing the YouTube videos included in IRA tweets.
The data used for this study include: 1)~Content data (tweets), 2)~User data (Twitter account information), and 3)~Moderation data (Accounts identified as IRA accounts).
Other research efforts focused on detecting actors involved in such campaigns. An example is the work from Luceri et al.~\cite{luceri2020detecting}, which focused on detecting state-sponsored trolls that were involved in the 2016 US elections using an inverse reinforcement learning approach. To achieve this, they used the following data: 
1)~\textit{Content Data} (tweets), 2)~\textit{User Data} (Twitter account information), and 3)~\textit{Moderation Data} (Accounts identified as IRA accounts).

Another method of recruiting user accounts for promoting the messages of a coordinated information campaign is employing social-media influencers. This strategy has been observed soon after Russia's invasion of Ukraine (``Russian lives matter'' campaign~\cite{richards_pro-russia_2022}) and has also been noted in the Romanian 2024 presidential elections~\cite{popoviciu_tiktok_2024}. We discuss more on this strategy in the context of hidden advertising in Section~\ref{sec:hidden-ads}. 

\textbf{Identifying Algorithmic Manipulation.} Algorithmic manipulation refers to the intentional exploitation of platform algorithms and systems to amplify/suppress specific narratives or distort user engagement~\cite{conti2024revealing}. 
Platform algorithms include recommendation systems (systems responsible for recommending content to users), algorithms for content search and ranking, and algorithms that identify trending topics based on user activity~\cite{elmas2021ephemeral}.
Nowadays, social media platforms increasingly rely on AI algorithms to enhance user experience, including recommendation systems to generate personalised feeds, recommend users to follow, etc.
Given the increasing reliance of social media platforms on AI-based algorithms for recommending content to users, there are several concerns with respect to how such systems can be exploited by malicious actors that aim to manipulate the public and promote/suppress specific narratives online.

Motivated by this, previous research focused on understanding these phenomena.
For instance, Conti et al.~\cite{conti2024revealing} conducted a quantitative study on suspicious changes in content visibility on Twitter, likely due to algorithmic interventions. To conduct their research, the authors used the following data: 
1)~\textit{Content Data} (tweets), 2)~\textit{User Data} (Twitter account information), 3)~\textit{Algorithmic Data} (content visibility as determined by the algorithm ), 4)~\textit{Temporal Data} (tweets over time), 5)~\textit{Engagement Data} (retweet information), and 6)~\textit{Moderation Data} (content factuality based on the sources).
Another example of algorithmic manipulation was shown by Elmas et al.~\cite{elmas2021ephemeral}, who demonstrated how the Twitter trending algorithm can be manipulated to promote specific political content in Turkey during the 2019 Istanbul elections. To demonstrate this kind of algorithmic manipulation, they used the following data: 1)~\textit{Content Data} (tweets), 2)~\textit{User Data} (Twitter account information), 3)~\textit{Algorithmic Data} (Trending topics), and 4)~\textit{Temporal Data} (algorithmic manipulation attacks over time).

\textbf{Summary of Data Requirements.} The overview below summarises the data points that could be relevant in the scientific investigation of platform manipulation based on prior literature and proposes examples of specific tasks which can use this data.

\begin{itemize}
\item \textbf{Content Data:} Text, images, videos, sound, transcriptions, hashtags, user mentions~\cite{allcott2017social,grinberg2019fake,zannettou2021won}. \textit{Example use case}: Analyse and detect posts that share false information during elections.
\item \textbf{User Data:} Account metadata, such as the account creation date, follower count, user demographics, bios, status, etc. \textit{Example use case}: Study the personas of accounts pertaining to state-sponsored actors~\cite{badawy2018analyzing,zannettou2019disinformation,luceri2020detecting}.
\item \textbf{Engagement Data:} Metrics such as likes, shares, comments, and views~\cite{allcott2017social,bovet2019influence,zannettou2021won}. \textit{Example use case}: Analysing user engagement with false information related to elections.
\item \textbf{Moderation Data:} Posts or content flagged, removed, or labelled by the platform or external sources (e.g., fact-checkers)~\cite{badawy2018analyzing,luceri2020detecting,zannettou2019disinformation,ferrara2020characterizing,shao2018spread}. Also, moderation data on the user level, e.g. whether a user account is suspected to be a bot or controlled by a state-sponsored actor. \textit{Example use case}: Studying the effectiveness of platform moderation interventions for posts sharing false information.
\item \textbf{Temporal Data:} Time-based activity logs, such as the engagement that a misinformation record receives over time~\cite{grinberg2019fake, zannettou2021won,conti2024revealing}. \textit{Example use case}: Analysing a timeline of the dissemination of a false narrative before an election. 
\item \textbf{Network Data:} Data about the social graph, such as user connections and interactions. \textit{Example use case}: Analysing influential nodes in disseminating false information~\cite{grinberg2019fake,badawy2018analyzing}.
\item \textbf{Algorithmic Data:} Data on algorithmic outputs and decisions (e.g., videos recommended to end-users), as well as the importance of various features in algorithmic outputs~\cite{elmas2021ephemeral,conti2024revealing}. \textit{Example use case}: Investigating changes in content visibility due to algorithmic manipulation.
\item \textbf{Geospatial Data:} Anonymised and aggregate location data~\cite{badawy2018analyzing,zannettou2019disinformation,bessi2016social,ferrara2017disinformation}. \textit{Example use case}: Identifying and analysing coordinated campaigns originating from specific regions.

\end{itemize}

\subsection{Political Influencer Marketing as Hidden Advertising}
\label{sec:hidden-ads}

The popularity of influencer marketing has been growing steadily from US\$35.1bn in 2024 to US\$52bn expected in 2028~\cite{statista_influencer_2024}. Its advantages, particularly the opportunity to embed advertising with organic content made by creators who establish parasocial relationships with their followers, make it an incredibly compelling marketing strategy -- even for political advertising. Candidates and political entities find it important to partake in the creator economy, and creators themselves are pressured to make more political content and engage with social justice questions~\cite{siu_how_2024}.

\textbf{Measuring the Prevalence of Undisclosed Ads.}
Undisclosed advertisements have been widely acknowledged as a major regulatory concern by various legislative bodies, including the European Commission (EC) drafting the DSA~\cite{duivenvoorde2023regulation}. Enforcement of the DSA in the realm of political advertising is of great significance since it could safeguard the free expression of diverse political ideologies and help maintain democratic integrity. In the context of the Romanian elections, political ideologies were disseminated through seemingly credible media channels, including social media influencers. When political advertising is not disclosed, audiences are less likely to recognise the content as advertisements (or other paid arrangements), leaving them vulnerable to manipulation~\cite{liu2024effects}. This lack of transparency allows certain political candidates to exploit misinformation and covert persuasion tactics, potentially swaying voter behaviour and gaining an unfair electoral advantage. Such practices not only undermine voter autonomy but also jeopardise the legitimacy of democratic processes. This issue aligns with broader concerns regarding systemic risks posed by hidden advertising as illegal content. Measuring the prevalence of hidden political advertisements is, therefore, a critical step in establishing whether they are a factor in the understanding of election-related systemic risks. By assessing how frequently and in what forms undisclosed political ads appear, regulators can gain valuable insights into the scope of the problem. This, in turn, can inform the development of targeted interventions to ensure elections remain fair, transparent, and free from manipulation, thereby upholding democratic values and fostering trust in political elections.

Measuring the prevalence of undisclosed ads first requires identifying which posts contain sponsored content. Existing sponsored content detection research has primarily focused on commercial content, framing the problem as a machine learning classification task. Many models adopt a semi-supervised approach, using disclosure hashtags or keywords (e.g., \#ad, \#sponsored) as weak labels to distinguish between sponsored and organic posts~\cite{kimDiscoveringUndisclosedPaid2021b,bertaglia2023closing}. Studies rely primarily on \textit{Content Data} (such as post captions, engagement metrics, and associated metadata) to train models~\cite{zareiCharacterisingDetectingSponsored2020,bertaglia2023closing,bertaglia2025influencer}. \citet{kimDiscoveringUndisclosedPaid2021b} incorporate \textit{Network Data} to model the relationship between influencers, brands, and (sponsored) posts and add more context to the machine learning models. Such approaches require additional \textit{User Data}, including account metadata (e.g., follower counts, account type). Other studies combine image and text data to improve model performance~\cite{sanchez-villegas-etal-2023-multimodal,villegas2021analyzing}. In this context, such multimodal models would benefit from additional \textit{Content Data}, including attributes inferred by the platforms (e.g., detected objects in videos, audio transcriptions, and visual markers of sponsorship). 

Beyond detecting undisclosed ads, measuring their prevalence at scale requires additional data, particularly about influencer accounts and their activity. For our election case study, researchers would need access to \textit{User Data} to identify political influencers, including lists of influencers, demographic attributes, account metadata (e.g., follower counts), and account types. \textit{Content Data}, such as text, images, videos, and inferred attributes (e.g. object detection) from influencer posts would be essential to detect undisclosed ads. \textit{Engagement Data}, including likes, shares, comments, and views, would help analyse implication patterns and detect coordinated behaviour. \textit{Network Data} would be important to map connections between influencers, political actors, and audience interactions. Finally, \textit{Moderation Data}, such as platform-inferred labels or flags related to sponsored or political content, could indicate existing detection mechanisms.

\textbf{Measuring the Engagement of Organic Content, Disclosed and Undisclosed Ads.}
A key reason behind the non-disclosure of advertisements is the widespread belief that revealing sponsorships harms audience engagement. Influencers, who simultaneously act as content creators and advertising platforms, grapple with the tension between protecting their long-term growth and maximising short-term profits. The assumption that disclosure reduces engagement incentivises some influencers to conceal sponsorships. Understanding the actual impact of advertisements and their (non)disclosure on engagement is therefore crucial for assessing the validity of this belief. Empirical studies have analysed the effects of disclosures and regulations on engagement, relying on \textit{Engagement Data} to compare audience reactions to disclosed versus undisclosed ads, as well as before and after specific regulations~\cite{ershovEffectsInfluencerAdvertising2020,bertaglia2025influencer}. However, effectively measuring these differences requires a longitudinal perspective; therefore, \textit{Temporal Data}, especially in the form of engagement metrics over time, is essential for understanding the persistence and impact of undisclosed advertising. Research can be designed to compare engagement levels across organic content and disclosed and undisclosed advertisements. If empirical findings refute the assumption that disclosure negatively affects engagement, this could pave the way for educational initiatives aimed at influencers, emphasising that transparency does not undermine engagement. Such efforts could, in turn, improve compliance rates with disclosure regulations.

\textbf{Assessing How Platforms Promote or Moderate Sponsored Content.} 
Another reason for the non-disclosure of advertisements stems from a belief among influencers that platforms deliberately restrict the visibility of advertising content~\cite{musiyiwa2023sponsorship,savolainen2022shadow,bishop2019managing}. This perception is linked to the inherent tension platforms face between increasing user traffic and maximising financial gains. While displaying more ads aligns with the interests of advertisers, it risks alienating users and discouraging them from engaging with the platform. Consequently, platforms are compelled to strike a balance between promoting and moderating advertisements to sustain their ecosystem. Due to this concern, an ideal choice for the influencers would be to hide the advertisement so that they can bypass the moderation process. To address this issue, it is crucial to understand how platforms’ content promotion and moderation mechanisms operate. Research could explore the factors influencing the visibility of advertising content, such as algorithmic preferences, user behaviour data and the balance between paid and organic content. By examining the biases inherent in these systems, such research could identify patterns that favour or limit certain types of content, providing valuable insights into how platforms manage the promotion and visibility of advertisements. In practice, investigating whether and how platforms moderate or amplify advertising content requires access to several key data types. In addition to the data points identified for the previous tasks, \textit{Algorithmic Data}, such as content recommended to users and features related to their ranking, would be essential for determining how sponsored content performs in the recommender system within the platform.

\textbf{Summary of Data Requirements.} The overview below summarises the data points that could be relevant in the scientific investigation of hidden advertising based on prior literature and proposes examples of specific tasks for the analysis of this data.

\begin{itemize}
    \item \textbf{Content Data:} All media content (text, images, videos, and sound), metadata (e.g., captions, timestamps, and hashtags), and inferred attributes (e.g., object detection and visual sponsorship markers) from influencer posts. These data points are fundamental for identifying advertising content~\cite{zareiCharacterisingDetectingSponsored2020,bertaglia2023closing,villegas2021analyzing,sanchez-villegas-etal-2023-multimodal}. \textit{Example use case}: Training machine learning models to detect undisclosed political ads.
    \item \textbf{User Data:} Metadata related to influencer accounts, including follower counts, account creation date, account type, and inferred demographics. Studying influencer engaging in political campaigns requires access to lists of content creators and their attributes~\cite{kimDiscoveringUndisclosedPaid2021b,bertaglia2025influencer}. \textit{Example use case}: Mapping Romanian influencers involved in political campaigns by analysing account demographics and activity.
    \item \textbf{Engagement Data:} Metrics such as likes, shares, comments, and views across influencer posts~\cite{ershovEffectsInfluencerAdvertising2020}. \textit{Example use case}: Analysing the impact of disclosures and regulation in engagement with political posts.
    \item \textbf{Moderation Data:} Information on platform-inferred labels, content flagging, removals, and downranking of sponsored and political content. This data would help determine whether automated moderation disproportionately affects disclosed versus undisclosed ads~\cite{musiyiwa2023sponsorship,savolainen2022shadow}. \textit{Example use case}: Assessing whether platform moderation algorithms label disclosed political ads differently from undisclosed ones.
    \item \textbf{Temporal Data:} Engagement metrics over time~\cite{ershovEffectsInfluencerAdvertising2020,bertaglia2025influencer}. \textit{Example use case}: Identify potential coordinated engagement with political content.
    \item \textbf{Network Data:} Interaction data capturing connections between influencers, brands, political entities, and audience members. Mapping these relationships helps detect coordinated campaigns and assessing the diffusion of undisclosed ads~\cite{kimDiscoveringUndisclosedPaid2021b}. \textit{Example use case}: Detecting coordinated influence operations by analysing how content spreads across influencer networks.
    \item \textbf{Algorithmic Data:} Includes recommendation system outputs, ranking metrics, and visibility logs for influencer content. Access to this data is essential to assess whether platform recommendation systems amplify or suppress disclosed advertisements compared to undisclosed content~\cite{savolainen2022shadow}. \textit{Example use case}: Analysing whether recommendation algorithms rank disclosed political ads differently from undisclosed ones.
\end{itemize}

\section{Data Availability as a Platform Governance Narrative}
\label{sec:platform-gov}

So far, we have focused on understanding what data would ideally be used to identify systemic risks. Knowing what data platforms have is essential for researchers to request data through Article 40 DSA. In their data access application under Article 8(6) DA, researchers are required to provide ``an explanation as to why the research project cannot be carried out with alternative existing means such as using data available through other sources'' -- an incentive for researchers not to overburden the data access framework. Therefore, the researcher needs to show that the data they need are not publicly available (e.g. in a research API). In addition, to improve their chances of getting their data access applications through, researchers should also ensure that the data they seek is actually internally available with platforms.    
\begin{figure*}
    \centering
    \includegraphics[width=1\linewidth]{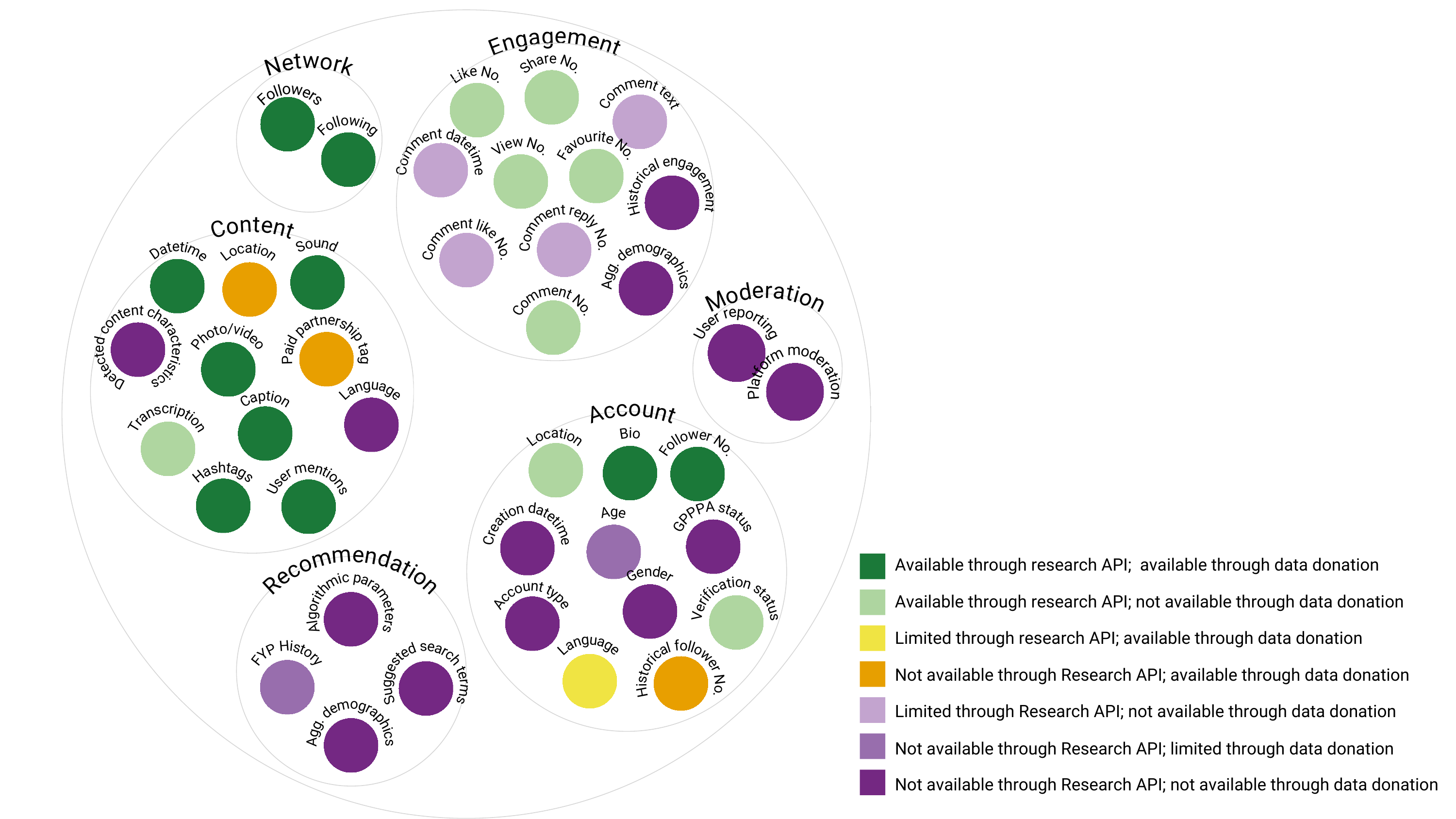}
    \caption{Availability of TikTok data for studying systemic risks in Romanian election interference incidents under existing access mechanisms, classified by data type and access method. Nested circles represent data categories with individual data points inside. Colours indicate accessibility, ranging from green (fully available via the Research API or data donations) to purple (entirely inaccessible to researchers), with intermediate shades representing partial access. The analysis highlights critical gaps in access, particularly in user attributes, algorithmic recommendations, and content moderation decisions, which limit researchers’ ability to investigate the systemic risks described in our scenarios.}
    \label{fig:data_overview}
\end{figure*}
How platforms collect, process and store personal data from users in exchange for access to platform services can be inferred in two ways: (i)~through \textit{platform documentation} on user-facing webpages constituting the contractual relationship between users and the platform; and (ii)~through \textit{visible data} in APIs and user-requested data (data donations). In this section, we analyse TikTok's documentation, its Research API, and data donations to identify discrepancies between what the platform says it collects and what it visibly collects. We then use this framework to analyse the gaps in the data needed to investigate platform manipulation and hidden advertising in the context of election interference, as outlined in the overviews of Sections 3.2 and 3.3.  


In TikTok's Privacy Centre,\footnote{\url{https://www.tiktok.com/privacy/overview/en.}} users can find a list of examples of ``information that we \textit{may} collect'' (emphasis added) and, in the Safety Centre,\footnote{\url{https://www.tiktok.com/safety/en/.}} a more detailed list of ``some'' of what is collected. Both user-facing webpages direct users to the Privacy Policy.\footnote{\url{https://www.tiktok.com/legal/page/eea/privacy-policy/en.}} While this addresses personal data collected from users, the TikTok Partner Privacy Policy covers information processed in relation to partner platforms that include TikTok for Business, TikTok Shop and TikTok Creator Marketplace. 

We focus on the user-facing Privacy Policy, which outlines ``what information we collect'', organised by three types: ``information you provide'', ``automatically collected information'', and ``information from other sources''. Under each category, the policy lists different types of information with specific data included as examples. While the nature of platform development necessitates flexibility in the policy, and users can understand categories of data, definitive data points only emerge in ``such as'' lists, impeding clarity of what data platforms have. For example, TikTok states, ``We infer your attributes (such as age-range and gender) and interests based on the information we have about you'' indicating the non-exhaustive nature of the wording. 

Drawing on the data mentioned in this Privacy Policy, we evaluate the availability of data requirements described in the case study. We find that not all the data required for studying the identified systemic risks are included. We then turn to other platform documentation, such as the \textit{Branded Content Policy}\footnote{\url{https://www.tiktok.com/legal/page/global/bc-policy/en.}} and \textit{Help Pages},\footnote{\url{https://support.tiktok.com/en/.}} to ascertain whether this is data that TikTok processes and so could be requested. For instance, algorithmic parameters and recommended search terms are generated by the platform and are not personal data that would need to be described in the privacy policy. Table~\ref{tab:data_points_source} lists the documentation source for each data point relevant to the case study. 

We then determine whether each data point is available through the Research API or data donations. The latter was obtained by the authors from their own TikTok research accounts. The API provides qualifying research access to certain public data, while data donations~\cite{van2021promises,zannettou2023leveraging} provide individual users with their personal data, which can then be donated to researchers. While other approaches, such as web scraping, are used for research purposes, we focus on these two mechanisms that comply with the terms of service of the platform and so would render a data access request through Article 40 unnecessary.

Using TikTok documentation for the API and data donations, which detail the exact data points available, we ensure that the form in which they are provided would be sufficient for the needs identified in the scenarios. We classify the availability of each data point as \textit{available}, \textit{not available}, or \textit{limited}. For example, we identify \textit{Comment} data and \textit{FYP history} as limited: the Research API limits the collection of comments to the top 1000 and does not specify what constitutes ``top''. Similarly, although viewing history is available through data donations, it includes not only content recommended to the user through the \textit{For You Page}, but also content viewed through other sources such as direct messages or the follower feed. Figure~\ref{fig:data_overview} illustrates the distribution of data availability. Each coloured circle represents a data point nested within a circle area showing the higher-level data category. The colours signify the level of availability; we consider the API as the most effective mechanism for data collection in our scenarios. For data donations, we would require access to either the accounts posting the election-related content or the users viewing such content, which are populations challenging to identify and recruit.

Our analysis demonstrates that the existing mechanisms for data access provide insufficient means to investigate the systemic risks described in our case study. Not only can we not access any data related to \textit{Recommendation} and \textit{Moderation}, but also, even in the categories in which some data are available, the other necessary corresponding data points are not, such as accounts' attributed demographics and account type, thereby rendering the research questions unanswerable. Problematically, inferred attributes about content and user demographics cannot be accessed, impeding our ability to contextualise data and understand targeting and recommendation practices. Therefore, researchers are left with a limited perspective that detaches content and engagement from the people targeted by coordinated campaigns.

\section{Conclusion: Reflections on DSA Data Access for the Research Community}
\label{sec:conclusions}

The implementation of DSA Article 40 is a significant milestone in the pursuit of greater platform transparency and accountability. By granting vetted researchers access to data, this mechanism has great potential to uncover systemic risks and to foster a more informed regulatory landscape. However, our analysis reveals critical legal interpretational and operational challenges that hinder the realisation of this potential, due to how bureaucratic and cumbersome Article 40 will prove to be in practice. Not only are systemic risks very vague to define and identify, but requesting data can lead to a standoff problem between platforms and researchers. In the information asymmetry defining digital industries such as social media, not having a general overview of the data availability might hurt researchers' chances of getting data access, either because the requested data may be public in one way or another, or because the data are invisible, and platforms may deny having them.

Our case study of the Romanian election interference highlights the complexity of defining and operationalising systemic risks under the DSA (Article 34(1)(c)). This case study reveals how two election-related systemic risk factors---platform manipulation and hidden advertising---ought to be studied and what data is needed for such scientific investigations based on established literature in the relevant computational fields. In our account of data requirements for investigating systemic risk in the case study, we establish that current mechanisms provide researchers with insufficient access to data, necessitating access through Article 40. To do this, we relied on terms of service and other platform documentation to justify the availability of requested data, providing evidence that TikTok gathers, infers, and processes specified data to counteract the issue of opacity in knowing what data exists. While we propose and demonstrate how this exercise can be effective for researchers in preparing applications, the issue of transparency persists, particularly concerning what invisible data TikTok may collect. But how platforms will indicate what data and data structures are available, and the extent to which this will capture the totality of data available to researchers, raises questions about how DSCs and the European Commission will further operationalise this provision to address the standoff referred to above. This is all the more relevant for cross-platform research, as our case study was limited to TikTok, but election interference is always prevalant beyond just one platform.

To address these challenges, we propose several general recommendations. First, clearer definitions and guidelines regarding the interpretation of systemic risks are essential to support researchers in framing their data requests effectively, particularly when national interpretations by DSCs can differ significantly. Second, improving transparency about the data platforms collect, infer, and process is crucial for resolving any ambiguity of researchers around data availability. Data inventories (Article 6(4) DA) provided by platforms could be a step in the right direction. The negotiation of what data should be mapped in such inventories could be further informed by additional case studies like the one discussed in this paper, which could identify, for different categories, the data gaps in publicly available or visible data. Finally, fostering collaboration between researchers, regulatory bodies, and platforms is necessary to ensure that the data access framework aligns with both academic and societal needs. The European Commission can play an important role in mediating exchanges of needs and interests to align expectations and ensure compliance with the transparency policy goals pursued by the DSA's data access regime. 

\section{Acknowledgments}
This research has been supported by funding from the ERC Starting Grant HUMANads (ERC-2021-StG No 101041824).

\bibliography{main}

\newpage
\section{Paper Checklist}
\begin{enumerate}

\item For most authors...
\begin{enumerate}
    \item  Would answering this research question advance science without violating social contracts, such as violating privacy norms, perpetuating unfair profiling, exacerbating the socio-economic divide, or implying disrespect to societies or cultures?
    \answerYes{Yes. This is a conceptual paper that analyses the DSA's Article 40 data-access mechanism and, via a case study, specifies what kinds of data would be needed to study systemic risks. We do not collect, process, or publish personal data, nor do we scrape platforms or build models that profile individuals or groups.}
  \item Do your main claims in the abstract and introduction accurately reflect the paper's contributions and scope?
    \answerYes{Yes. The abstract and introduction accurately state our contributions.}
   \item Do you clarify how the proposed methodological approach is appropriate for the claims made? 
    \answerYes{Yes. We clarify that this is a conceptual and methodological paper: we derive data needs from prior literature on systemic risks, map them onto the Romanian election case study, and compare these needs with what TikTok documentation, the Research API, and data donations provide. This approach is appropriate for the claims we make, as our contribution is to outline methodological requirements and gaps, not to produce new empirical findings.}
   \item Do you clarify what are possible artifacts in the data used, given population-specific distributions?
    \answerNA{NA. This paper does not use empirical datasets.} 
  \item Did you describe the limitations of your work?
    \answerYes{Yes. We discuss the limitations of our work, including the interpretational uncertainty of systemic risks under the DSA, the lack of transparency about what data platforms actually collect, and the insufficiency of currently available mechanisms (APIs and data donations) to study election-related systemic risks. We also acknowledge that our case study focuses only on TikTok, not covering the cross-platform nature of systemic risks.}
  \item Did you discuss any potential negative societal impacts of your work?
    \answerNA{NA. As a conceptual paper, we do not conduct empirical analysis or release tools or datasets that could have direct negative societal impacts.}
      \item Did you discuss any potential misuse of your work?
    \answerNA{NA. The paper does not produce datasets, tools, or empirical models that could be misused. Our contribution is limited to conceptual analysis and a case study framing of data needs under the DSA.}
    \item Did you describe steps taken to prevent or mitigate potential negative outcomes of the research, such as data and model documentation, data anonymization, responsible release, access control, and the reproducibility of findings?
    \answerNA{NA.}
  \item Have you read the ethics review guidelines and ensured that your paper conforms to them?
    \answerYes{Yes.}
\end{enumerate}

\item Additionally, if your study involves hypotheses testing...
\begin{enumerate}
  \item Did you clearly state the assumptions underlying all theoretical results?
    \answerNA{NA.}
  \item Have you provided justifications for all theoretical results?
    \answerNA{NA.}
  \item Did you discuss competing hypotheses or theories that might challenge or complement your theoretical results?
    \answerNA{NA.}
  \item Have you considered alternative mechanisms or explanations that might account for the same outcomes observed in your study?
    \answerNA{NA.}
  \item Did you address potential biases or limitations in your theoretical framework?
    \answerNA{NA.}
  \item Have you related your theoretical results to the existing literature in social science?
    \answerNA{NA.}
  \item Did you discuss the implications of your theoretical results for policy, practice, or further research in the social science domain?
    \answerNA{NA.}
\end{enumerate}

\item Additionally, if you are including theoretical proofs...
\begin{enumerate}
  \item Did you state the full set of assumptions of all theoretical results?
    \answerNA{NA.}
	\item Did you include complete proofs of all theoretical results?
    \answerNA{NA.}
\end{enumerate}

\item Additionally, if you ran machine learning experiments...
\begin{enumerate}
  \item Did you include the code, data, and instructions needed to reproduce the main experimental results (either in the supplemental material or as a URL)?
    \answerNA{NA.}
  \item Did you specify all the training details (e.g., data splits, hyperparameters, how they were chosen)?
    \answerNA{NA.}
     \item Did you report error bars (e.g., with respect to the random seed after running experiments multiple times)?
    \answerNA{NA.}
	\item Did you include the total amount of compute and the type of resources used (e.g., type of GPUs, internal cluster, or cloud provider)?
    \answerNA{NA.}
     \item Do you justify how the proposed evaluation is sufficient and appropriate to the claims made? 
    \answerNA{NA.}
     \item Do you discuss what is ``the cost`` of misclassification and fault (in)tolerance?
    \answerNA{NA.}
  
\end{enumerate}

\item Additionally, if you are using existing assets (e.g., code, data, models) or curating/releasing new assets, \textbf{without compromising anonymity}...
\begin{enumerate}
  \item If your work uses existing assets, did you cite the creators?
    \answerNA{NA.}
  \item Did you mention the license of the assets?
    \answerNA{NA.}
  \item Did you include any new assets in the supplemental material or as a URL?
    \answerNA{NA.}
  \item Did you discuss whether and how consent was obtained from people whose data you're using/curating?
    \answerNA{NA.}
  \item Did you discuss whether the data you are using/curating contains personally identifiable information or offensive content?
    \answerNA{NA.}
\item If you are curating or releasing new datasets, did you discuss how you intend to make your datasets FAIR?
\answerNA{NA.}
\item If you are curating or releasing new datasets, did you create a Datasheet for the Dataset? 
\answerNA{NA.}
\end{enumerate}

\item Additionally, if you used crowdsourcing or conducted research with human subjects, \textbf{without compromising anonymity}...
\begin{enumerate}
  \item Did you include the full text of instructions given to participants and screenshots?
    \answerNA{NA.}
  \item Did you describe any potential participant risks, with mentions of Institutional Review Board (IRB) approvals?
    \answerNA{NA.}
  \item Did you include the estimated hourly wage paid to participants and the total amount spent on participant compensation?
    \answerNA{NA.}
   \item Did you discuss how data is stored, shared, and deidentified?
   \answerNA{NA.}
\end{enumerate}

\end{enumerate}

\appendix
\begin{table*}[ht]
\section{Appendix A}
\caption{Availability of TikTok data for studying systemic risks in Romanian election interference incidents, classified by data category and access mechanism. Each row maps individual data points to their corresponding documentation sources and indicates their accessibility through the Research API and data donations.}
\label{tab:data_points_source}
\begin{tabular}{@{}lllll@{}}
\toprule
\textbf{Category} & \textbf{Data}                    & \textbf{Documentation}               & \textbf{Research API} & \textbf{Data Donation} \\ \midrule
Content           & Photo/ video                     & Privacy Policy                       & Yes                                         & Yes                                       \\
Content           & Caption                          & Privacy Policy                       & Yes                                         & Yes                                       \\
Content           & Create datetime                  & Privacy Policy                       & Yes                                         & Yes                                       \\
Content           & Create location                  & Privacy Policy                       & No                                          & Yes                                       \\
Content           & Hashtags                         & Privacy Policy                       & Yes                                         & Yes                                       \\
Content           & User mentions                    & Privacy Policy                       & Yes                                         & Yes                                       \\
Content           & Sound                            & Privacy Policy                       & Yes                                         & Yes                                       \\
Content           & Transcription                    & Privacy Policy                       & Yes                                         & No                                        \\
Content           & Language                         & Help Centre\tablefootnote{\url{https://support.tiktok.com/en/live-gifts-wallet/tiktok-live/comments-on-tiktok-live/.}} & No                                          & No                                        \\
Content           & Paid partnership tag             & Branded Content Policy               & No                                          & Yes                                       \\
Content           & Detected content characteristics & Privacy Policy                       & No                                          & No                                        \\
Moderation        & Reporting by users               & Privacy Policy                       & No                                          & No                                        \\
Moderation        & Moderation by platform           & Privacy Policy                       & No                                          & No                                        \\
Account           & Age                              & Privacy Policy                       & No                                          & Limited                                   \\
Account           & Gender                           & Privacy Policy                       & No                                          & No                                        \\
Account           & Language                         & Privacy Policy                       & Limited                                     & Yes                                       \\
Account           & Bio                              & Privacy Policy                       & Yes                                         & Yes                                       \\
Account           & Location                         & Privacy Policy                       & Yes                                         & No                                        \\
Account           & Follower count                   & Privacy Policy                       & Yes                                         & Yes                                       \\
Account           & Historical follower count        & Privacy Policy                       & No                                          & Yes                                       \\
Account           & Creation datetime                & Privacy Policy                       & No                                          & No                                        \\
Account           & Account type                     & Privacy Policy                       & No                                          & No                                        \\
Account           & GPPPA status                     & Help Centre\tablefootnote{\url{https://support.tiktok.com/en/using-tiktok/growing-your-audience/government-politician-and-political-party-accounts/.}}          & No                                          & No                                        \\
Account           & Verification status              & Help Centre\tablefootnote{\url{https://support.tiktok.com/en/using-tiktok/growing-your-audience/how-to-tell-if-an-account-is-verified-on-tiktok/.}}    & Yes                                         & No                                        \\
Engagement        & Like count                       & Privacy Policy                       & Yes                                         & No                                        \\
Engagement        & Share count                      & Privacy Policy                       & Yes                                         & No                                        \\
Engagement        & View count                       & Privacy Policy                       & Yes                                         & No                                        \\
Engagement        & Favourite count                  & Privacy Policy                       & Yes                                         & No                                        \\
Engagement        & Comment count                    & Privacy Policy                       & Yes                                         & No                                        \\
Engagement        & Comment text                     & Privacy Policy                       & Limited                                     & No                                        \\
Engagement        & Comment creation datetime        & Privacy Policy                       & Limited                                     & No                                        \\
Engagement        & Comment like count               & Privacy Policy                       & Limited                                     & No                                        \\
Engagement        & Comment reply count              & Privacy Policy                       & Limited                                     & No                                        \\
Engagement        & Historical engagement metrics    & Privacy Policy                       & No                                          & No                                        \\
Engagement        & Aggregate demographics           & Privacy Policy                       & No                                          & No                                        \\
Network           & Username of followers            & Privacy Policy                       & Yes                                         & Yes                                       \\
Network           & Username of following            & Privacy Policy                       & Yes                                         & Yes                                       \\
Recommendation    & FYP history                      & Privacy Policy                       & No                                          & Limited                                   \\
Recommendation    & Algorithmic parameters           & Help Centre\tablefootnote{\url{https://support.tiktok.com/en/using-tiktok/exploring-videos/for-you/.}}                 & No                                          & No                                        \\
Recommendation    & Aggregate demographics           & Privacy Policy                       & No                                          & No                                        \\
Recommendation    & Suggested search terms           & Help Centre\tablefootnote{\url{https://support.tiktok.com/en/using-tiktok/exploring-videos/how-tiktok-recommends-content/.}}          & No                                          & No                                        \\ \bottomrule
\end{tabular}
\end{table*}

\end{document}